# Towards a standardised strategy to collect and distribute application software artifacts

Thomas Laurenson, Stephen MacDonell, Hank Wolfe

*Department of Information Science, School of Business, University of Otago.*
*thomas@thomaslaurenson.com, [stephen.macdonell, hank.wolfe]@otago.ac.nz*

## Abstract

*Reference sets contain known content that are used to identify relevant or filter irrelevant content. Application profiles are a type of reference set that contain digital artifacts associated with application software. An application profile can be compared against a target data set to identify relevant evidence of application usage in a variety of investigation scenarios. The research objective is to design and implement a standardised strategy to collect and distribute application software artifacts using application profiles. An advanced technique for creating application profiles was designed using a formalised differential analysis strategy. The design was implemented in a live differential forensic analysis tool, LiveDiff, to automate and simplify data collection. A storage mechanism was designed based on a previously standardised forensic data abstraction. The design was implemented in a new data abstraction, Application Profile XML (APXML), to provide storage, distribution and automated processing of collected artifacts.*

**Keywords:** Differential analysis, application profiles, reverse engineering, application software, Digital Forensic XML

## 1. INTRODUCTION

Application software are the computer programs that perform specific end-user tasks (e.g., web browsers, word processors and image editors). Forensic analysis of application software aids digital event reconstruction by revealing digital artifacts (e.g., file system entries and system configuration information). These artifacts are a robust source of evidence regarding application software usage in specific scenarios.

Reference sets contain known content usually represented by metadata, which are compared to an investigation target to identify relevant matches or to perform data reduction. For example, a reference set for a malicious tool can be compared against a perpetrator's hard drive to determine the presence of anti-forensic or hacking tools. Reference sets of application software have a variety of different names: application profile, footprint, fingerprint and signature. The term application profile is used in this paper.

Authoring application profiles involves system-level reverse engineering. Past researchers have reverse engineered a wide variety of applications to aid digital investigations. For example, the instant messaging application Digsby (Yasin & Abulaish, 2013), the cloud storage client Dropbox (Quick & Choo, 2013) and anti-forensic tools (Geiger & Cranor, 2006). In these studies the following method was carried out: 1) Manual analysis using a variety of reverse engineering techniques and tools; 2) Documentation of the analysis method and findings; and 3) Sharing of knowledge (usually via academic publication). This technique poses a variety of challenges for both researchers and practitioners.

1) **Reverse engineering techniques lack standardisation:** Researchers lack a systematic approach compounded by the fact that there are no standard set of tools, no tool automation and results that are unable to be shared (Garfinkel, 2010). Present research is a stand-alone endeavour with minimal technology advances.

2) **Challenges incorporating multiple evidence sources:** Reference sets are primarily comprised of metadata that represent data files (Roussev, 2010). However, most application software stores configuration information in the Windows Registry (Morgan, 2008). There are currently no methods to store and process multiple evidence sources in a single application profile.

3) **Application profile generation time:** Modern applications are regularly patched and updated, which means that maintaining a reference set for every software version is becoming less feasible (Roussev, 2011). Research has attempted to solve this using small block forensics (Garfinkel et al., 2010) and similarity digests (Roussev & Quates, 2012). A different solution would be to improve the speed and simplicity of data collection to enable rapid application profile creation. This would also increase application code coverage.

A standardised and automated approach would address these problems. Firstly, an automated live data collection method would streamline application profile generation. Secondly, a standardised data abstraction would facilitate the storage, distribution and automated processing of application software artifacts.

This paper outlines background material covering the theory and frameworks implemented in the proposed system design. A formalised process to identify application software artifacts is presented covering potential evidence sources, a data collection method, data collection procedure and a differential analysis strategy. A data abstraction suitable for application profile distribution is designed that specifies a structure, classification scheme, inclusion of pertinent metadata properties and standardisation using an Extensible Markup Language (XML) schema. Finally, a conclusion including future research areas is presented.

## 2. BACKGROUND: THEORY AND FRAMEWORKS

Previous researchers have advanced reference sets to improve data abstraction functionality, developed reverse engineering techniques and incorporated applicable evidence sources, which will now be discussed.

### Digital Forensics XML

Digital Forensics XML (DFXML) is an XML language designed to represent forensic information. Garfinkel (2009) developed the `fiwalk` tool to automate disk image processing by extracting file system metadata and populating a DFXML document. A Python API (`dfxml.py`) provides investigators with an object orientated approach to write simple automated scripts (Garfinkel, 2012). DFXML was extended by Nelson (2012) to include Windows Registry entries. Nelson et al. (2014) then formalised the DFXML language using an XML schema to provide document validation. A revised Python API (`Objects.py`) was implemented which provides mutative object properties and DFXML schema adherence.

### Differential Analysis

Differential forensic analysis is a standardised strategy to reverse engineer application software. It compares and reports the differences between two objects. Garfinkel et al. (2012) formalised a general differential forensic analysis strategy which reports the differences between any two kinds of digital artifacts; for example, two hard drives. The general strategy is:

$$A \xrightarrow{R} B$$

Garfinkel et al., (2012) stated that "if A and B are disk images and the examiner is evaluating the installation footprint of a new application, then R might be a list of files and registry entries that are created or changed". The output from differential forensic analysis can be used to construct an application profile by determining system-level changes using differential analysis. Garfinkel et al. (2012) released two differential analysis tools: 1) `idifference.py` compares two disk images and reports the file system differences; and 2) `rdifference.py` compares two Windows Registry hive files and reports the differences. Both tools use DFXML to perform post-mortem differencing. In contrast, `Regshot` is a live differential analysis tool that determines file system and Registry changes by comparing snapshots on a running system (Carvey, 2011).

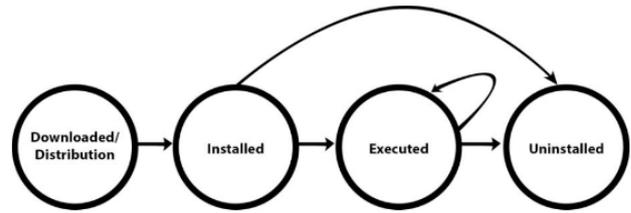

Figure 1: High-level overview of the application life cycle (Source: Figure adapted from Davis et al., (2006).

### Application Software Life Cycle

Each application has a life cycle that follows a chronological path including phases such as installation, execution, and uninstallation. During each phase of the application life cycle, digital artifacts are created, modified and/or removed. For example, when installing an application, various folders, files, and configuration settings are created. When uninstalling an application, these are removed but residual information may remain. Figure 1 displays a high-level overview of the application life cycle.

## 3. DATA GENERATION: AUTHORING APPLICATION PROFILES

This section outlines an overview of the proposed system to identify application software artifacts on a live operating system by implementing differential analysis. The sources of application software evidence, a data collection method and a novel technique to include efficient data file hashing is specified. A formalised differencing strategy and a scalable procedure for application life cycle recreation are also outlined.

### System Design Overview

The `Regshot` tool provides an efficient live system snapshot and comparison implementation but lacks sufficient reporting detail and file hashing capability. In contrast, the `idifference.py` and `rdifference.py` tools provide exceptionally detailed metadata reports but lack efficiency due to post- mortem analysis (Garfinkel et al., 2012). Combining both approaches would simplify application profile generation and to achieve this the system would require the following functionality:

1) A portable Windows tool to execute on a live system
2) Support to process file system and Windows Registry entries
3) Automated interface to ease application profile generation
4) Inclusion of cryptographic hashing for data files
5) Output to a standardised XML data abstraction

**Application Software Evidence Sources**
Application software creates, modifies and/or removes a variety of digital artifacts on an operating system. When investigating application software usage on a Microsoft Windows operating system the majority of digital artifacts of forensic interest are file system and system configuration information. Therefore, the following evidence sources should be included in an application profile: 1) File system entries (directories and data files); and 2) Windows Registry entries (keys and values).

**Data Collection Method**
The system design requirements specify support for a portable Windows tool to process file system and Registry entries. A new live differential forensic analysis tool, named `LiveDiff`, was authored base on the `Regshot` software. The `fileshot.c` and `regshot.c` source code files provide the functionality to snapshot the local file system and Registry and perform differencing (Regshot, 2015). The specified `Regshot` source code files were used as the foundation for the `LiveDiff` tool. However, numerous modifications and additional code were essential to implement the required functionality.

File system data collection is achieved by performing a snapshot of the system drive (usually C:\) whereas Registry data collection is accomplished by performing a snapshot of the HKEY_LOCAL_MACHINE (HKLM) and HKEY_USERS (HKU) Registry hives. This incorporates the SAM, SECURITY, SOFTWARE, SYSTEM and NTUSER.DAT hive files. Each snapshot is stored in a C data structure (SNAPSHOT). Table 1 displays the implemented data structures used to store digital artifact information. In addition to the listed properties in Table 1 each structure retains a pointer to associated father, brother and/or sub structure.

After performing two snapshots, a comparison is made to determine the system changes that have occurred. To accomplish this, a differential analysis strategy is required.

**Inclusion of Cryptographic Hashing for Data Files**
Roussev (2010) states that cryptographic file hashing is commonly used in digital investigations to identify data files that are exactly the same. Therefore, an application profile requires that data files must have an accompanying hash value to aid data file matching against a target data set. However, hashing every data file on a target system is computational inefficient, especially when performing live data collection.

A novel method was designed and implemented to perform selected file hashing. Before data collection is performed, an initial system snapshot is collected and used to create a blacklist of known files. The blacklist is stored in memory using a prefix tree (trie) data structure which is populated using the full path of all data files from the initial snapshot. The prefix tree provides an ordered tree data structure to provide fast string indexing. When performing subsequent system snapshots (i.e., data collection), the file path of data files are queried against the prefix tree, if no match is found the data file is hashed using the Secure Hash Algorithm version 1 (SHA-1). This implementation results in only new files of forensic interest being hashed.

Table 1: Overview of data structures used for different application software artifacts.

| Digital artifact | Structure name | Properties |
|---|---|---|
| Data file | FILECONTENT | File name, size, write time, access time, hash value and attribute |
| Directory | FILECONTENT | Directory name, size, write time, access time and attribute |
| Registry key | KEYCONTENT | Key name, modified time |
| Registry value | VALUECONTENT | Value name, type, data and data size |

**Differential Analysis Strategy**
The proposed differencing algorithm is implemented based on the general differential forensic analysis strategy specified by Garfinkel et al. (2012). The differencing algorithm can be expressed as:

$$Snapshot1 \xrightarrow{R} Snapshot2$$

*Snapshot1* is the system state before an application life cycle phase is conducted (e.g., application installation). *Snapshot2* is the system state after an application life cycle phase is conducted. The two snapshots are then compared (*R*) and the created, changed, modified and/or removed digital artifacts are reported. Figure 2 displays the algorithm used for file system entry correlation between snapshots (*FC* refers to FILECONTENT structures).

The differencing algorithm for Registry entries (keys and values) follows a very similar differential analysis strategy. However, differencing of Registry values is performed in an embedded loop after two matching Registry keys are discovered. All entries deemed new, changed, modified or deleted by the differencing algorithm are added to a data structure (RESULTS) which can be later processed and reported to the investigator.

```
1: procedure COMPAREFILES
2:    for each FILECONTENT (FC1) in Snapshot1 do
3:       for each FILECONTENT (FC2) in Snapshot2 do
4:          if (FC2 has previously been matched) then skip FC2
5:          end if
6:          if (FC1 type and name does not equal FC2 type and name) then skip FC2
7:          end if
8:          if FC1 is a file then
9:             if (all FC1 properties equal all FC2 properties) then FC2 is matched
10:            end if
11:            if (FC1 write time does not equal FC2 write time) then FC2 is changed
12:            end if
13:            if (FC1 size or hash does not equal FC2 size or hash) then FC2 is modified
14:            end if
15:         end if
16:         if FC1 is a directory then
17:            if (all FC1 properties equal all FC2 properties) then FC2 is matched
18:            end if
19:         end if
20:      end for
21:      if FC1 is not matched then FC1 must be deleted
22:      end if
23:   end for
24:   for each FILECONTENT (FC2) in Snapshot2 do
25:      if FC2 has not been matched then FC2 must be new
26:      end if
27:   end for
28: end procedure
```

Figure 2: Differential analysis strategy for file system entries.

**Data Collection Procedure**
`LiveDiff` was intentionally implemented as a console application to reduce user interaction and provided an automated data collection process for faster and simpler tool operation. Figure 3 displays the method used to achieve automated application profile generation using the `LiveDiff` tool.

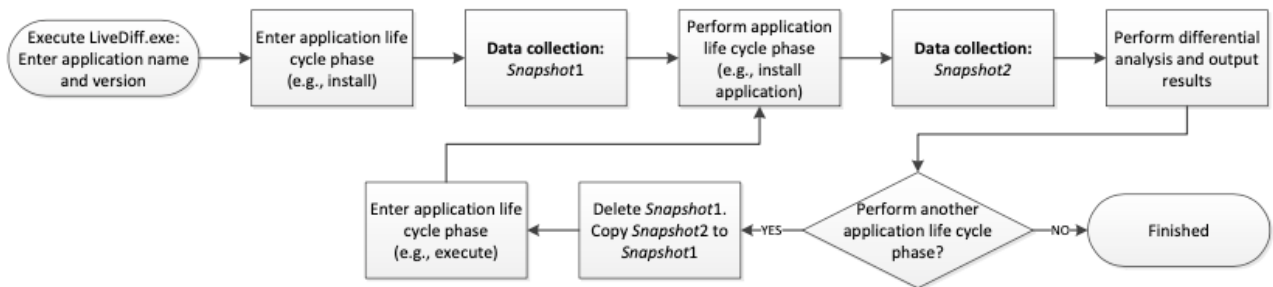

Figure 3: High-level overview of the LiveDiff data collection procedure.

The data collection procedure is a simple automated procedure that requires minimal user interaction. The user is prompted to enter the application name and version number. For each life cycle phase the user must: 1) Enter the life cycle state; 2) Press enter to collect Snapshot1; 3) Perform the application life cycle phase (e.g., install the application); 4) Press enter to collect Snapshot2. Differencing is performed and results reported by appending populated DFXML objects to an output file. The user can select to continue profile generation and perform another life cycle phase, or finish the scanning process. If an additional life cycle phase is requested by the user, Snapshot1 is deleted and Snapshot2 is copied to Snapshot1. This increases application profile generation speed by removing the requirement to recollect the first snapshot. All results obtained from snapshot comparison are populated into a specifically designed XML data abstraction that is discussed in the following section.

## 4. DATA ABSTRACTION: DISTRIBUTING APPLICATION PROFILES

A standardised and effective data abstraction would aid in creating and distributing application profiles. The data abstraction requires the functionality to store, distribute and automate processing of a variety of digital artifact types and provide sufficient information to classify application software artifacts.

**Data Abstraction Structure**

A suitable data abstraction has the following requirements:

1) Conforms to existing digital forensic requirements (e.g., evidence integrity)
2) Functionality to document file system and Registry entries
3) Standardised, extensible and open design

DFXML was selected as it conforms to the specified requirements. Thus, a new data abstraction, Application Profile XML (APXML), was designed based on the DFXML standardised data abstraction. Figure 4 displays a skeleton example of the proposed APXML structure.

A well-formed XML document must contain one root XML element (tag) (Yergeau et al., 2004). The APXML root element is defined using an `apxml` tag. Namespace attributes are recommended by the XML specification for uniquely named XML elements, therefore, the root element specifies an XML schema that was created based on the specifications of this research. The following Uniform

```
<?xml version='1.0' encoding='UTF-16' ?>
<apxml version="'1.0.0'"
  xmlns="https://github.com/thomaslaurenson/apxml_schema"
  xmlns:dc="http://purl.org/dc/elements/1.1/"
  xmlns:xsi="http://www.w3.org/2001/XMLSchema-instance"
  xmlns:delta="http://www.forensicswiki.org/wiki/Forensic_Disk_Differencing">
  <metadata/>
  <creator/>
  <install>
      <!-- DFXML FileObjects -->
      <!-- RegXML CellObjects -->
  </install>
  <execute>
      <!-- DFXML FileObjects -->
      <!-- RegXML CellObjects -->
  </execute>
  <uninstall>
      <!-- DFXML FileObjects -->
      <!-- RegXML CellObjects -->
  </uninstall>
</apxml>
```

Figure 4: Example of the Application Profile XML (APXML) structure.

Table 2: Overview of the metadata properties for different digital artifact types stored in an Application Profile XML document. DFXML Objects.py naming conventions are used.

| File System | | Windows Registry | |
|---|---|---|---|
| **Directory** | **File** | **Key** | **Value** |
| filename | filename | cellpath | cellpath |
| meta_type | meta_type | name_type | name_type |
| alloc_name | sha1 | alloc | data_type |
| alloc_inode | alloc_name | | data |
| | alloc_inode | | alloc |

Resource Identifier (URI) specifies the APXML namespace: `https://github.com/thomaslaurenson/apxml_schema` The schema provides compliance to the unique element naming conventions in an APXML document. A number of additional XML namespaces are required as an APXML document includes DFXML `FileObject` entries, RegXML `CellObject` entries, DFXML delta annotations (to describe `FileObject` and `CellObject` differencing states) and XML Dublin Core to annotate the XML document.

Similar to the DFXML standard, an APXML document has both metadata and creator elements to document case provenance. The metadata element documents additional information about an application profile including the profiled application name and version while the creator element documents information pertaining to the tool that authored the APXML document and the environment it was executed in. The creator element implemented in the APXML structure is taken from the DFXML standard (version 1.0). The remainder of the APXML structure categorises digital artifacts based on the application life cycle phases.

**Digital Artifact Classification**
A key component of the APXML structure is the classification of digital artifacts to provide application life cycle information. Each digital artifact is represented by a specific DFXML object. File system entries are populated in `FileObjects` and Registry entries are populated in `CellObjects`. An APXML document classifies each object using a naming convention to describe life cycle phase association. The APXML structure outlines four classifications based on the application life cycle: 1) Install; 2) Execute; 3) Uninstall; and 4) Reboot. Due to the open and extensible design the APXML structure can be extended to include additional life cycle phases and different naming conventions. Digital artifact classification provides an investigator with additional information regarding application software usage. For example, installing an application is a different scenario to that of installing then executing the software for a specific task. Both scenarios provide evidence that can be used to determine what tasks a perpetrator conducted with an application.

**Digital Artifact Metadata Properties**
DFXML stores detailed metadata about digital artifacts. However, not all metadata is required in an application profile. This is because only certain metadata properties would aid digital artifact correlation against a target data set. For example, the full file system path and corresponding hash value can be used to perform digital artifact detection. In contrast, partition information and file timestamps would not aid digital artifact correlation as these values would differ between target systems. Table 2 displays the required metadata properties for directories, files, Registry keys and values stored by APXML documents.
Each of the metadata properties store different information dependent on the digital artifact type. Table 3 displays the various metadata properties with an accompanying description and examples.

**Standardising the Application Profile XML Structure**
XML document validation is an important process that ensures correct data structure for tool production or consumption. The DFXML language was formalised via implementation of an XML schema and validation can be achieved using the `xmllint` utility (Nelson et al., 2014). This research adopts the same approach. An XML schema (`apxml.xsd`) was created to validate APXML documents to ensure correct production and consumption of APXML documents. This provides researchers and practitioners with the capability to distribute reverse engineering results with assurance of document validity and usability.

Table 3: Summary of the metadata property types used in an Application Profile XML (APXML) document with a description and examples.

| Property | Description | Example |
|---|---|---|
| filename | Full file system path | Program Files/TrueCrypt/TrueCrypt.exe |
| meta_type | File system entry type | 1 = file |
| | | 2 = directory |
| sha1 | SHA-1 hash value | 7689d038c76bd1df695d295c026961e50e4a62ea |
| alloc_name | File allocation status | 1 = allocated |
| | | 0 = unallocated |
| alloc_inode | Metadata allocation status | 1 = allocated |
| | | 0 = unallocated |
| cellpath | Full Registry entry path | HKLM/Software/Classes/AppID/TrueCrypt.exe |
| name_type | Registry entry type | k = Registry key |
| | | v = Registry value |
| data_type | Registry value data type | REG_SZ = Null terminated string |
| | | REG_DWORD = 32-bit number |
| data | Registry value data | @C:\Program Files\TrueCrypt\TrueCrypt.exe |
| alloc | Cell allocation status | 1 = allocated |
| | | 0 = unallocated |

## 5. CONCLUSION

This research contributed towards a standardised strategy to collect and distribute application software artifacts. A new live differential analysis tool was authored, `LiveDiff`, which simplifies and accelerates the generation of application profiles using an automated process. An advanced data abstraction, Application Profile XML (APXML), was designed which incorporates multiple evidence sources into a single document using an accepted forensic data abstraction format. The data abstraction was standardised using an XML schema. The output of the research culminates in a system designed based on accepted digital forensic requirements that can aid researchers and practitioners to reverse engineer, store, distribute and automate processing of application software artifacts.

Forensic analysis of application software is still an active research area that requires additional investigation. The research conducted would benefit from a practical evaluation covering tool effectiveness and efficiency. Additional evidence sources could be included in the APXML document including volatile memory information, document signatures and network traffic signatures. Inclusion of different hashing algorithms (block hashing and similarity digests) could advance application profile functionality to detect similar but not exact copies of digital artifacts. All of these future research areas would require expansion of the DFXML standard to document the specified evidence sources and hashing algorithms.

There are a variety of other research areas involving generating application profiles. Filtering irrelevant digital artifacts to exclude operating system noise from the data collection phase has yet to be investigated. Alternative methods for data collection could improve profile generation techniques. For example, performing differential analysis using virtual machine snapshots taken before and after application software life cycle phases.

**Resource Availability**
The resources supporting this research have been made publicly available to encourage future research and development. The `LiveDiff` tool and APXML schema (`apxml.xsd`) is available from the authors GitHub repositories: https://github.com/thomaslaurenson